\documentclass[twocolumn,preprintnumbers,amsmath,amssymb,nofootinbib,superscriptaddress]{revtex4}

\usepackage{amsmath}
\usepackage{amsfonts}
\usepackage{amssymb}
\usepackage{graphicx, rotating}
\usepackage{epstopdf}
\usepackage{epsfig}
\usepackage{latexsym}
\usepackage{graphicx}
\usepackage{color}
\usepackage{amsmath,amssymb}
\usepackage{slashed}
\usepackage{hyperref}
\hypersetup{colorlinks, citecolor=bluscuro, linkcolor=black, urlcolor=bluscuro}
\definecolor{rossos}{cmyk}{0,1,1,0.55}
\definecolor{bluscuro}{rgb}{0.15, 0.2, .85}
\definecolor{bluchiaro}{cmyk}{1,.3,0.,0.1}

\newcommand{\eq}[1]{Eq.~(\ref{#1})}

\newcommand{\be}{\begin{equation}}
\newcommand{\ee}{\end{equation}}
\newcommand{\bea}{\begin{eqnarray}}
\newcommand{\eea}{\end{eqnarray}}

\newcommand{\GeV}{\,\mathrm{GeV}}

\begin{document}
%\begin{center}
%{\Large \bf Peering through the Higgs Portal \\ with Monojets}
%\vspace{2 cm}
%
%{\large Francesco Riva\,$^a$ and  Alfredo Urbano\,$^b$}
%
%\vspace{1cm}
%
%\centerline{$^a${\it  IFAE, Universitat Aut\`onoma de Barcelona, 08193
%Bellaterra, Barcelona, SPAIN}}
%\vspace{0.2cm}
%\centerline{$^b${\it  Laboratoire de Physique Th\'eorique de l'\'Ecole Normale Sup\'erieure,}}
%\centerline{{\it  24 rue Lhomond, F-75231 Paris, FRANCE}}
%
%\end{center}
%\vspace{.8cm}

%\begin{abstract}
%
%
% \end{abstract}
%\newpage

\title{Higgs discovery: the beginning or the end of natural EWSB?}
\date{\today}

\author{Marc Montull}
\email{mmontull@ifae.es}
\affiliation{IFAE, Universitat Aut{\`o}noma de Barcelona,
08193 Bellaterra, Barcelona, Spain}

\author{Francesco Riva}
\email{friva@ifae.es}
\affiliation{IFAE, Universitat Aut{\`o}noma de Barcelona,
08193 Bellaterra, Barcelona, Spain}

\begin{abstract}
We use global fits to analyze the most recent Higgs data from ATLAS, CMS and Tevatron and compare the Standard Model (SM) prediction with natural extensions of the SM. In particular we study wide classes of composite Higgs models based on different coset structures (leading at low energy to different Higgs sectors including extra singlets and Higgs doublets) and different coupling structures of the elementary fermions to the strong sector. We point out in what situations the composite models could improve (or worsen) the fit to the data and compare with similar trends in the MSSM.
\end{abstract}
\maketitle

A particle consistent with the SM Higgs boson has been discovered: 5$\sigma$ deviations have been observed both by  CMS \cite{CERNTalkCMS} and ATLAS \cite{CERNTalkATLAS}, in the combination of $\gamma\gamma$ and $ZZ$ channels. Whether this is the beginning or the end of an era of investigation of natural realizations of the electroweak scale, it is not yet clear.

In this note we perform a global analysis of how compatible this excesses are with the SM expectation of a Higgs boson at $m_h\approx125 \GeV$, using the most recent data from ATLAS, CMS and Tevatron. It is then interesting to analyze different natural theories beyond the standard model (SM), to see whether they would be preferred or disfavored by the present trend of data, and to understand towards which direction the parameter space of such theories is more likely to shrink.
We first  turn our attention to a large variety of composite Higgs models \cite{Agashe:2004rs}, highlighting which features in these models tend to improve/worsen the fit to the data.  In section \ref{sec:54}, we begin with the minimal composite Higgs models (MCHMs), with coset structure SO(5)/SO(4), but with generic structures of fermion couplings to the strong sector (and hence to the Higgs boson) \cite{Agashe:2004rs,Contino:2006qr,Pomarol:2012qf}.

Larger coset structures can have strikingly different phenomenology. For instance in SO(6)/SO(5), which includes a CP-odd singlet besides the Higgs doublet in its light spectrum \cite{Gripaios:2009pe}, the singlet couples to $\gamma\gamma$ through the Wess-Zumino-Witten term; this can enhance $h\rightarrow \gamma \gamma$  if the Higgs and the singlet mix with each other. We show this in section \ref{sec:65} and also study the case where, due to extra accidental symmetries, the singlet can  manifest itself through invisible Higgs decays. The possibility of mixing raises the question of whether it will be possible  at the LHC to understand if the particle we observe is really the one whose VEV gives mass to the gauge bosons; we try to answer this question by singling out the exclusive channels (the ones with vector boson fusion (VBF) and associated production cuts) that carry mostly this information.

We then turn to the last class of composite models, based on the coset structure $SO(6)/SO(4)\times SO(2)$ which delivers an effective two Higgs doublets model (THDM) at low energy \cite{Mrazek:2011iu}. We compare the predictions of this composite version of the THDM with that of the MSSM in section \ref{sec:642}.

Global fits of Higgs data in the context of composite Higgs models, but limited to the MCHM4 \cite{Agashe:2004rs} and MCHM5 \cite{Contino:2006qr} models, have already appeared in Refs.~\cite{Espinosa:2010vn,Carmi:2012yp,Azatov:2012bz,Espinosa:2012ir,Ellis:2012rx}.

\section{The Data}
\label{sec:DATA}

We assume the existence of a unique Higgs-like state with couplings to the SM-gauge bosons and fermions
\begin{equation}\label{equation1}
c_t\equiv \frac{y_t}{y_t^{SM}},\,\,c_b\equiv \frac{y_b}{y_b^{SM}},\,\,c_\tau\equiv \frac{y_t}{y_\tau^{SM}},\,\,a\equiv\frac{g_{hVV}}{g_{hVV}^{SM}},
\end{equation}
where we use the SM couplings as reference values and assume $g_{hVV}\equiv g_{hWW}=g_{hZZ}$. We take that the probability density functions (PDFs) provided by the experiments can be approximated by Gaussian distributions, and we use the theoretical prediction for the ratio \cite{Azatov:2012bz},
\begin{equation}
\mu^i=\frac{\sum_p\sigma_p(a,c_t,c_b,c_\tau)\zeta_p^i}{\sum_p\sigma_p^{SM}\zeta_p^i}\frac{BR_i(a,c_t,c_b,c_\tau)}{BR_i^{SM}},
\end{equation}
for each channel $i$ with production crossections $\sigma_p$ and cut efficiencies $\zeta_p^i$ (which we take to be independent from the parameters $a$ and $c_{t,b,\tau}$; the values of $\zeta_p^i$ are discussed in the Appendix).
We sum theoretical \cite{crossx} and experimental errors in quadrature (both errors are first symmetrized by average in quadrature and when negative error bars are not provided by the experimental collaborations, we have assumed symmetric distributions around the mean value). We summarize the data used in table~\ref{tableChannels}.\footnote{When data has been provided only in the combination of the 7 TeV and 8 TeV runs, we extract the information about the 8 TeV run assuming that the PDFs corresponding to the combined data can be written as the product of uncorrelated  PDFs i.e. PDF$_{7+8}$=PDF$_7$PDF$_8$.}
\begin{table}[b]
\caption{\emph{CMS, ATLAS  and  Tevatron data for the most sensitive channels. The cuts are classified as inclusive (I), associated production (A), vector boson fusion (VBF) or else ($\gamma\gamma_X$), see Appendix.
$\hat{\mu}^{1.96,7,8}$ denote the best fits for the 1.96 TeV Tevatron, and the 7,8,7+8 TeV  LHC data. }}
\begin{center}
\begin{tabular}{|c|c|c|c|c|}
\hline
CMS  & Cuts & $\hat{\mu}^7$ & $\hat{\mu}^8$ &$\hat{\mu}^{7+8}$  \\ [2 pt]
\hline
 $\gamma \gamma_0$ \cite{CMSPhotons} & $\gamma\gamma_X$ & $3.1^{+1.9}_{-1.8}$  & $1.5^{+1.3}_{-1.3}$& -\\ [2 pt]
%\hline
 $\gamma \gamma_1$ \cite{CMSPhotons}& $\gamma\gamma_X$ & $0.6^{+1.0}_{-0.9}$ & $1.5^{+1.1}_{-1.1}$  & -\\ [2 pt]
%\aline
 $\gamma \gamma_2$ \cite{CMSPhotons} & $\gamma\gamma_X$& $ 0.7^{+1.2}_{-1.2} $& $1.0^{+1.2}_{-1.2}$& -\\ [2 pt]
%\hline
 $\gamma \gamma_3$ \cite{CMSPhotons} & $\gamma\gamma_X$ & $1.5^{+1.6}_{-1.6}$ & $3.8^{+1.8}_{-1.8}$& -\\ [2 pt]
%\hline
 $\gamma \gamma_{jj}$ \cite{CMSPhotons} & $\gamma\gamma_X$
   & $4.2^{+2}_{-2}$ & $\begin{array}{lr}L:&-0.6^{+2.0}_{-2.0}\\T:&1.3^{+1.6}_{-1.6}\end{array}$ & -\\ [2 pt]
%\hline
 $\tau \tau$ \cite{Collaboration:2012tx008,CERNTalkCMS}  & I & $0.6^{+1.1}_{-1.3}$ & -& $-0.2^{+0.7}_{-0.7}$\\
%\hline
 $ b b$  \cite{Collaboration:2012tx008,CERNTalkCMS}& A & $1.2^{+2.1}_{-1.9}$& -&$0.1^{+0.8}_{-0.7}$ \\
%\hline
$WW_{0j}$   \cite{CERNTalkCMS} & G & $0.1^{+0.6}_{-0.6}$ & $1.3^{+0.8}_{-0.6}$ & - \\
%\hline
$WW_{1j}$  \cite{CERNTalkCMS}  & G & $1.7^{+1.2}_{-1.0}$ & $0.0^{+0.8}_{-0.8}$ & - \\
%\hline
$WW_{2j}$  \cite{CERNTalkCMS} & VBF & $0.0^{+1.3}_{-1.3}$ & $1.3^{+1.7}_{-1.3}$& - \\
%\hline
$ZZ$ \cite{Collaboration:2012tx008,CERNTalkCMS}& I & $0.6^{+1.0}_{-0.6}$& - & $0.7^{+0.5}_{-0.4}$\\
\hline
\hline
$\begin{array}{c}\textrm{ATLAS}\\125\GeV\end{array}$ & Cuts & $\hat{\mu}^7$  & $\hat{\mu}^{8}$&$\hat{\mu}^{7+8}$     \\ [2 pt]
\hline
$ \gamma \gamma$ \cite{ATLAS:2012tx019,Atlasphoton}& I & $1.6^{+0.8}_{-0.7}$& $0.9^{+0.5}_{-0.7}$& -\\
%\hline
 $\tau \tau$  \cite{ATLAS:2012tx019}& I & $0.2^{+1.7}_{-1.8}$& -& - \\
%\hline
$ b b$ \cite{ATLAS:2012tx019}& A & $0.5^{+2.1}_{-2.0}$ & -& -\\
%\hline
$WW$ \cite{ATLAS:2012tx019}& I & $0.6^{+0.7}_{-0.7}$ & - & -\\
%\hline
$ZZ$ \cite{ATLAS:2012tx019,CERNTalkATLAS}& I & $1.4^{+1.3}_{-0.8}$& - & $1.3^{+0.6}_{-0.6}$\\
\hline
\hline
CDF/D0  & Cuts & $\hat{\mu}^{1.96}$ & -  & -\\
\hline
$\gamma \gamma$  \cite{:2012cn}& I & $3.6^{+3.0}_{-2.5}$& - & -\\
%\hline
$bb$ \cite{:2012cn}& A & $2.0^{+0.7}_{-0.6}$  & -& - \\
%\hline
$WW$ \cite{:2012cn}& I & $0.3^{+1.2}_{-0.3}$& -& - \\
\hline
\hline
$\begin{array}{c}\textrm{ATLAS}\\126.5\GeV\end{array}$  & Cuts & $\hat{\mu}^7$  & $\hat{\mu}^{8}$&$\hat{\mu}^{7+8}$     \\ [2 pt]
\hline
$ \gamma \gamma$  \cite{ATLAS:2012tx019,Atlasphoton}& I & $2.0^{+0.8}_{-0.7}$& $1.7^{+0.7}_{-0.6}$& -\\
%\hline
 $\tau \tau$   \cite{ATLAS:2012tx019}& I & $0.3^{+1.7}_{-1.8}$ & -& - \\
%\hline
$ b b$  \cite{ATLAS:2012tx019}& A & $0.5^{+2.2}_{-2.2}$& -& -\\
%\hline
$WW$  \cite{ATLAS:2012tx019}& I & $0.5^{+0.6}_{-0.6}$ & - & -\\
%\hline
$ZZ$  \cite{ATLAS:2012tx019,CERNTalkATLAS}& I & $1.1^{+1.0}_{-0.7}$& - & $1.0^{+0.6}_{-0.5}$\\
\hline
\end{tabular}\end{center}
\label{tableChannels}
\end{table}%

ATLAS finds that the peak of the combined signal strength's best fit is at $m_h=126.5$ GeV, which is within experimental error from $m_h=125.3$ GeV, where the peak of CMS occurs, so one can assume them to belong to the same resonance. We perform the statistical analysis taking the values at $m_h=125$ GeV for CMS and Tevatron and $m_h=126.5$ GeV for ATLAS. For comparison we also study the case where $m_h=125$ GeV is assumed for all experiments, this appears in the plots as dashed lines.

Taking the ATLAS data at 126.5(125) GeV we obtain for the SM
\begin{equation}
\chi^2_{SM} = 26.36(23.9),
\end{equation}
which for $N=33$ independent channels corresponds to $\chi^2/N\approx 0.8(0.7)$.
For the parameters mentioned above, in the case with $c\equiv c_t=c_b=c_\tau$, the best fit  ($\chi^2$=19.4(19.9) which, for $33$ channels and 2 variables, $N=33-2=31$ corresponds to $\chi^2/N\approx 0.6$) occurs in
\begin{equation}
c=-0.69(-0.61),\,\, a=0.86(0.83),
\end{equation}
while another local probability maximum ($\chi^2$= 22.7(20.7), $\chi^2/N\approx0.7$) occurs for positive $c$ at
\begin{equation}
c=0.69(0.68),\,\, a=1.02(0.98).
\end{equation}

\begin{figure}[h!]
\centering
   \includegraphics[scale=0.68]{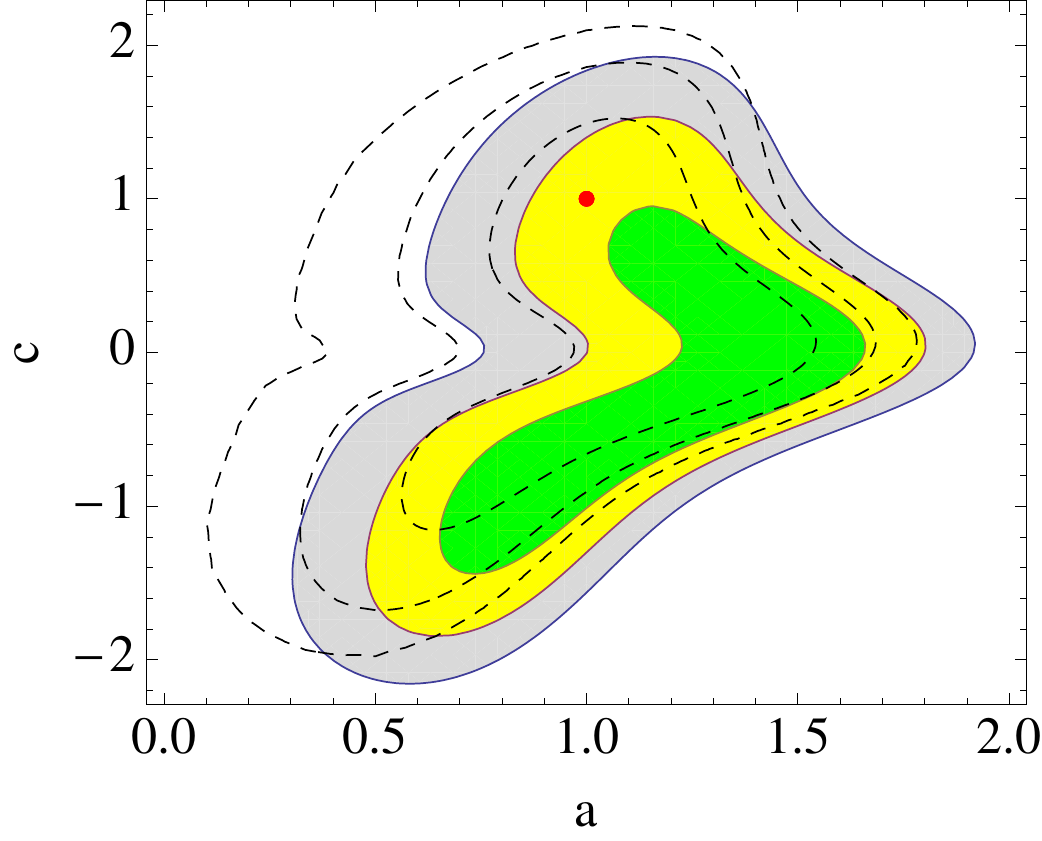}
   \includegraphics[scale=0.665]{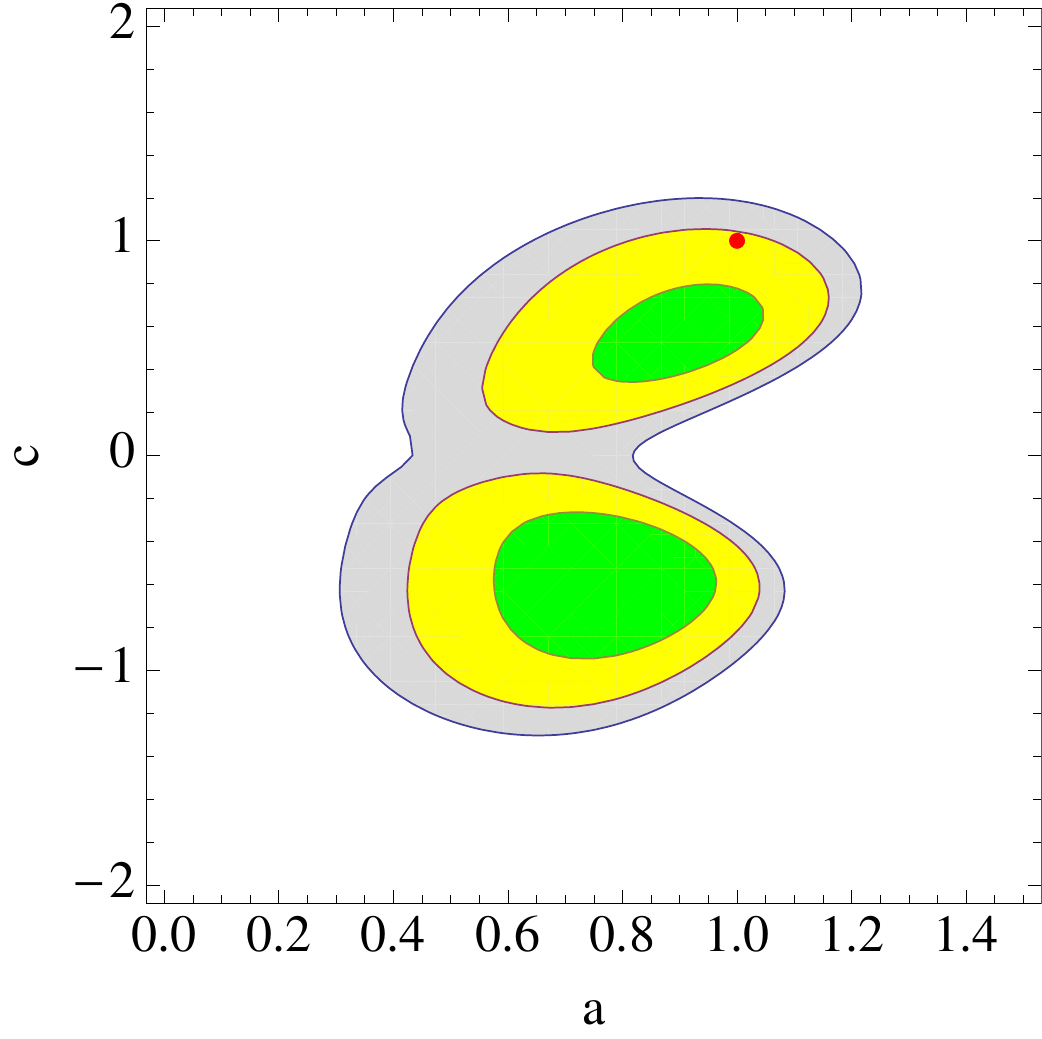}
 \caption{\footnotesize \emph{In green, yellow and gray, the 68\%,95\%,99\% C.L. contours for the parameters $a$ and $c$ with the most recent data (table~\ref{tableChannels}). {\it Upper plot:} ATLAS with data taken at $m_h=126.5$ GeV  (dashed contours correspond to data taken at $m_h=125$GeV). {\it Lower plot:}CMS with data taken at $m_h=125$GeV. A flat prior $a\in[0,3]$, $c\in[-3,3]$ is used. }}
 \label{figureAC}
\end{figure}

In fig.~\ref{figureAC} we show the 68\%,95\% and 99\% C.L. contours for the parameters $a$ and $c$ from a global fit of data from ATLAS and CMS. C.L. regions are found by finding the isocontour of $P(x,y)={\rm const}$ such that $\int dx \, dy \, P(x,y) \pi(x,y)= 0.99, 0.95, 0.68$,
where $x,y$ are any of the parameters $a, c_t, c_b, c_\tau$ shown in the specific plot, $\pi(x,y)$ is a flat prior and $P(x,y)=\prod_i {\rm PDF_i(x,y)}$ (product over all the channels where PDFs are given).

%%%%%%%%%%%%%%%%%%%%%%%%%%%%%%%%%%%%%%%%%%%%%%%%%%%%%%%%%%%%
%%%%%%%%%%%%%%%%%%%%%%%%%%%%%%%%%%%%%%%%%%%%%%%%%%%%%%%%%%%%
%%%%%%%%%%%%%%%%%%%%%%%%%%%%%%%%%%%%%%%%%%%%%%%%%%%%%%%%%%%%
\section{Composite Higgs Models}
%%%%%%%%%%%%%%%%%%%%%%%%%%%%%%%%%%%%%%%%%%%%%%%%%%%%%%%%%%%%
%%%%%%%%%%%%%%%%%%%%%%%%%%%%%%%%%%%%%%%%%%%%%%%%%%%%%%%%%%%%
%%%%%%%%%%%%%%%%%%%%%%%%%%%%%%%%%%%%%%%%%%%%%%%%%%%%%%%%%%%%

As shown above, the best fits occur for  modified couplings of the Higgs boson to the SM fermions and gauge bosons. This is a typical features of Composite Higgs models \cite{Giudice:2007fh}: for instance, due to the Pseudo Nambu-Goldstone boson (PNGB) nature of  the Higgs, the couplings between $h$ and the $W,Z$ gauge bosons are modified as
\begin{equation}\label{aMCHCM}
a=\sqrt{1-\xi},
\end{equation}
where $\xi\equiv v^2/f^2$, $f$ being the analogue of the pion decay constant and $v=246$ GeV is the vacuum expectation value (VEV) of the Higgs field. Interestingly, on the one hand $\xi\ll1$  from constraints coming from electroweak precision data (EWPD); on the other hand $\xi$ is a measure of fine-tuning in these models\footnote{\label{footnotetuning}The loop-induced potential for the PNGBs is a function of $\sin v/f$ and, without any fine-tuned cancellation, would naturally induce $v\approx f$ or $v=0$.} and is expected to be sizable.

%%%%%%%%%%%%%%%%%%%%%%%%%%%%%%%%%%%%%%%%%%%%%%%%%%%%%%%%%%%%
%%%%%%%%%%%%%%%%%%%%%%%%%%%%%%%%%%%%%%%%%%%%%%%%%%%%%%%%%%%%
\section{SO(5)/SO(4) and different fermion couplings}
\label{sec:54}
%%%%%%%%%%%%%%%%%%%%%%%%%%%%%%%%%%%%%%%%%%%%%%%%%%%%%%%%%%%%

While the strong sector alone is SO(5) symmetric, the couplings of elementary fermions to the strong sector break this symmetry, since  the SM fermions do not fill complete SO(5) multiplets. We can parametrize these couplings as spurions which transform both under the SM-gauge group and under some representation $\bf r$ of SO(5) (the well known minimal models MCHM4 \cite{Agashe:2004rs} and MCHM5 \cite{Contino:2006qr} correspond to  ${\bf r}=4$ and ${\bf r}=5$, respectively). Depending on the size of {\bf r}, the coupling of $h$ to fermions $f$ might deviate from the SM as
\cite{Pomarol:2012qf}:
\be
c_f=\frac{1+2m-(1+2m+n)\xi}{\sqrt{1-\xi}}\, ,
\label{deformation}
\ee
where $m,n$ are positive integers which depend on ${\bf r}$.  The specific cases with $m=n=0$ or $m=0$, $n=1$  correspond to the MCHM4 (with $c=\sqrt{1-\xi}$) and MCHM5 (with $c=(1-2\xi)/\sqrt{1-\xi}$), where all fermions share the same coupling structure.
Models with $m\neq0$ have deviations w.r.t. the SM of order unity (in the direction $c>1$), even in the limit $\xi\rightarrow 0$ and we shall not consider them any further.

\begin{figure}[h!]
\centering
   \includegraphics[scale=0.7]{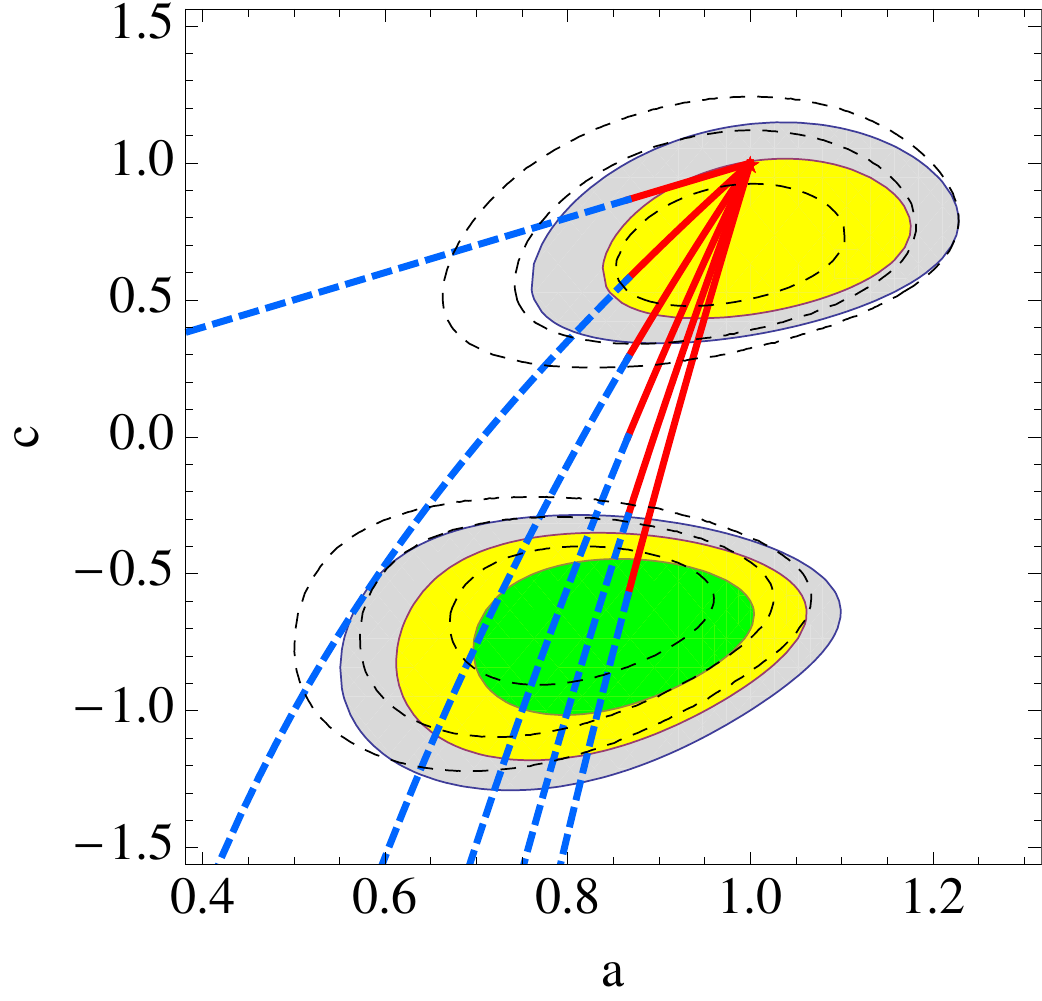}
 \caption{\footnotesize \emph{Global fit for the parameters $a$ and $c$, obtained combining CMS and Tevatron for $m_h=125$ GeV and ATLAS for $m_h=126.5$ (dashed circles use ATLAS at $m_h=125$ GeV); colors  and priors as in fig.~1. The lines denote predictions  of a generic MCHM; different curves  correspond to different values of  $n=0,...,5$ in \eq{deformation} ($m=0$),
  going downwards ($n=0,1$ correspond to the MCHM4 and MCHM5). The red part of the curves is for $0<\xi<0.25$ and  the blue dashed for $0.25<\xi<1$.}}
 \label{figureghff}
\end{figure}

In the specific case with $c\equiv c_t=c_b=c_\tau$, the effects of \eq {aMCHCM} and \eq{deformation} can be well described in the $(a,c)$ plane. We compare this theoretical expectation, for $m=0$ and $n=0,...,5$, with the best fit from the combined results of ATLAS (at $m_h=126.5$ GeV) and CMS ($m_h=125$ GeV), for the parameters ($a,c$)  in fig.~\ref{figureghff} (the dashed contours show the same fit taking the ATLAS data at $m_h=125$ GeV). We assume that no states, beside the SM ones, contribute via loop-effects to the $hgg$ and $h\gamma\gamma$ vertices.

Interestingly, representations leading to large $n\gtrsim 4$ can fit well the data also in the region with $c<0$, where the rate $h\rightarrow  \gamma \gamma$ is enhanced, due to a positive interference between $W$ and $t$ loops in the $h\gamma\gamma$ vertex (the fact that it is possible to have order 1 changes in this coupling, from modification of order $O(v^2/f^2)\ll 1$ is due to the large $n\gtrsim 4$ enhancement). To our knowledge, explicit models of this type do not exist yet in the literature ($n=4$ would appear in models where the spurions connecting SM fields and the strong sector transform as an irreducible representation ${\bf r}\in 5\otimes5\otimes5\otimes5$ of SO(5)) and it would be interesting to see if realistic models can be built.

As a final example, we consider the possibility of coupling top and down-type ($b$ and $\tau$) fermions in different ways to the strong sector (models of this type have been proposed, for instance, in refs.\cite{Mrazek:2011iu,Frigerio:2012uc}). We show examples of this as dots in the $c_t,|c_b|$ plane in fig.~\ref{figureTHDM1} with $c_t=\sqrt{1-\xi}$ and $c_b=c_\tau=(1-2\xi)/\sqrt{1-\xi}$ (black dot) and with $c_t=(1-2\xi)/\sqrt{1-\xi}$ and $c_b=c_\tau=\sqrt{1-\xi}$ (gray dot). Fig.~\ref{figureTHDM1} shows slices of constant $a$: for this reason these models, which map into a curve in the 3D $(a,c_b,c_t)$-space, appear as dots in the figure.
The asymmetric couplings do not  improve the fit to the data, which shows a preference for the region $c_b\approx c_t$.

%%%%%%%%%%%%%%%%%%%%%%%%%%%%%%%%%%%%%%%%%%%%%%%%%%%%%%%%%%%%
%%%%%%%%%%%%%%%%%%%%%%%%%%%%%%%%%%%%%%%%%%%%%%%%%%%%%%%%%%%%
\section{SO(6)/SO(5) and the role of the extra singlet}
\label{sec:65}
%%%%%%%%%%%%%%%%%%%%%%%%%%%%%%%%%%%%%%%%%%%%%%%%%%%%%%%%%%%%
%%%%%%%%%%%%%%%%%%%%%%%%%%%%%%%%%%%%%%%%%%%%%%%%%%%%%%%%%%%%

The light spectrum of models based on the $SO(6)/SO(5)$ coset structure contains, beside the Higgs doublet, an extra CP-odd scalar $\eta$ \cite{Gripaios:2009pe}. In the absence of extra symmetries, the scalar potential is generic and
 the Higgs scalar mixes with the singlet: this is the most important conceptual difference w.r.t. the $SO(5)/SO(4)$ models as far as $h$-phenomenology is concerned. The mass-eigenstate basis reads
\begin{equation}
\left(
\begin{array}{c}
 \eta^\prime \\
  h^\prime
\end{array}
\right)=\left(
\begin{array}{cc}
 \cos\alpha& \sin\alpha \\
  -\sin\alpha&\cos\alpha
\end{array}
\right)\left(\begin{array}{c}
 \eta \\
 h
\end{array}\right),\label{Mixing}
\end{equation}
where we take the mixing angle $\alpha$ a free parameter. The couplings of $\eta$ (being a gauge singlet) to the SM fermions and gauge bosons arise only at the non-renormalizable level and are suppressed by powers of  $v/f$. As a consequence,  mixing implies that all the couplings of the physical Higgs will be generally  suppressed by  $\cos\alpha$.

Now, it is important to notice that in composite Higgs models no enhancement is generally expected for $h\rightarrow\gamma\gamma$ \cite{Falkowski:2007hz}(see however  refs.~\cite{Azatov:2011qy,Bellazzini:2012tv} where exceptions are discussed). The reason is that effective operators, mediated by heavy states, of the form
\begin{equation}
c_B \frac{H^\dagger H}{f^2} B_{\mu\nu}B^{\mu\nu},\,\,c_W \frac{H^\dagger H}{f^2} W_{\mu\nu}W^{\mu\nu},
\end{equation}
 are generically supposed to be small by NDA arguments \cite{Giudice:2007fh}: they break the shift symmetry of the PNGB Higgs and their coefficient should be suppressed by powers of a weak coupling over strong coupling.
The coset $SO(6)/SO(5)$, however, admits a Wess-Zumino-Witten (WZW) term in its effective Lagrangian, corresponding to a quantum anomaly of the UV theory. At leading order  in $1/f$ (and without including the effect of possible couplings of $\eta$ to the SM fermions) this includes
\begin{gather}\label{anomcoupling}
\mathcal{L} \subset \frac{\eta} {32 \pi^2 f} (n_B B_{\mu\nu}\tilde{B}^{\mu\nu} + n_W W_a^{\mu\nu}\tilde{W}^a_{\mu\nu}).
\end{gather}

\begin{figure}[h]
\centering
   \includegraphics[scale=0.6]{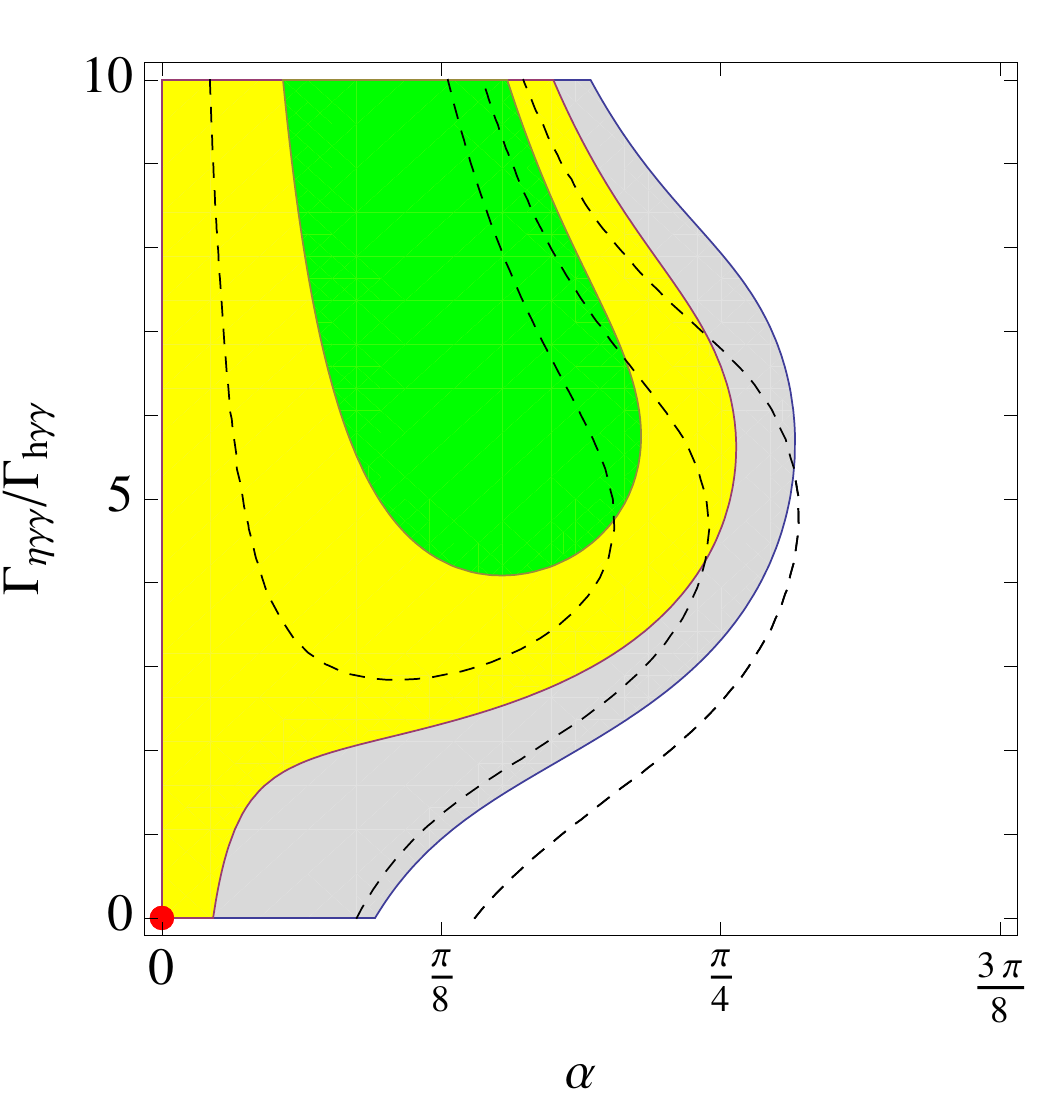}
 \caption{\footnotesize \emph{Tee best global fit as a function of $\alpha$ (the mixing between $h$ and the singlet $\eta$) and the ratio of the widths $ \Gamma_{\eta\gamma \gamma}/ \Gamma_{h\gamma \gamma}$; here $\xi=0.25$ and  $m_h=125$ GeV are fixed and  a flat prior in $\alpha\in[-\pi/2,\pi/2]$ and $\Gamma_{\eta\gamma\gamma}/\Gamma_{h\gamma\gamma}\in[0,10]$ is used. Colors as in fig.~1.}}
 \label{figuregammas}
\end{figure}

Therefore, in the $SO(6)/SO(5)$ models, while all  cross-section times branching ratios are reduced by $\cos^2\alpha$, the WZW term induces a coupling between $\eta$ and photons and can enhance the BR for photons (only\footnote{We ignore the contribution of \eq{anomcoupling} to the $hVV$ vertex ($V=W,Z$), contribution which will always be smaller then the renormalizable one.}):
\begin{equation}
\Gamma_{h\gamma \gamma}\rightarrow \Gamma_{h\gamma \gamma}(\cos^2\alpha+\sin^2\alpha\frac{ \Gamma_{\eta\gamma \gamma}}{ \Gamma_{h\gamma \gamma}}),
\label{hgammas}
\end{equation}
where $\Gamma_{\eta\gamma \gamma}$ is the width of $\eta\rightarrow\gamma\gamma$ evaluated at $m_\eta=m_h$ (in what follows $ \Gamma_{\eta\gamma \gamma}$ will be taken as a free parameter). Assuming that the SM fermions couple universally to the strong sector through spurions in the representation ${\bf r}=6$ of $SO(6)$, the actual couplings of $h$ to fermions/vectors become
\begin{equation}
c=\cos{\alpha}\frac{1-2\xi}{\sqrt{1-\xi}},\quad a=\cos\alpha \sqrt{1-\xi}.
\end{equation}
We show this situation in fig.~\ref{figuregammas}, where it can be seen that a sizable width of $\eta\rightarrow\gamma\gamma$ and a non-vanishing, but small, mixing angle can improve the fit w.r.t. the SM. Interestingly, as for $\pi_0\rightarrow\gamma\gamma$, the width of $\eta\rightarrow\gamma\gamma$,  being non-renormalized, could carry information about the structure of the UV theory: for instance a large number of colours $N\gg N_c$ in the new strong sector could lead naturally to $ \Gamma_{\eta\gamma \gamma}\gg \Gamma_{h\gamma \gamma}$.

\begin{figure}[]
  \begin{minipage}{0.2\textwidth}
   \centering
   \includegraphics[width=3.8cm]{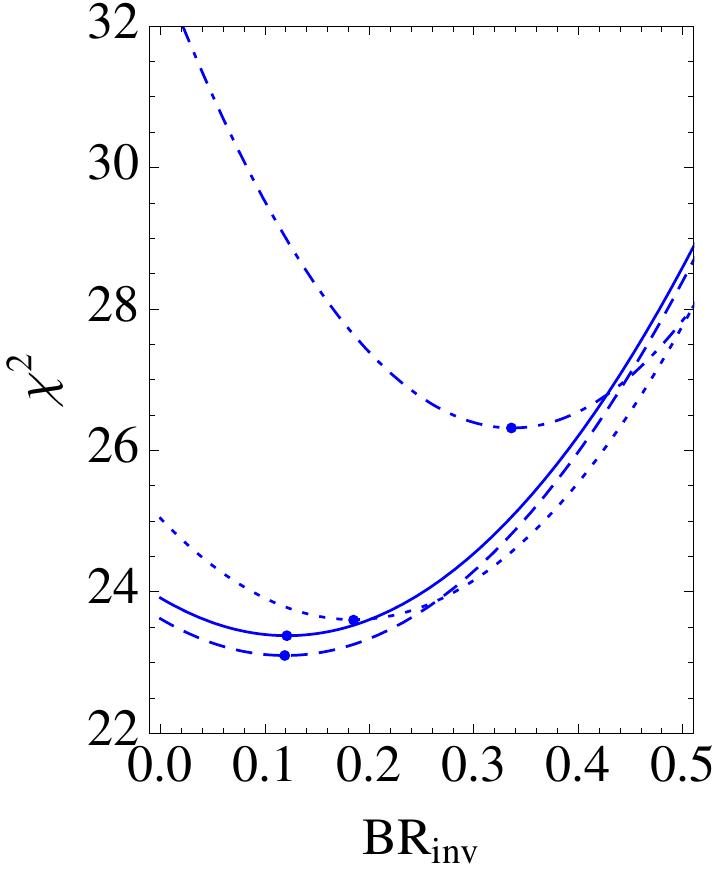}
    \end{minipage}\hspace{0.5 cm}
   \begin{minipage}{0.2\textwidth}
    \centering
    \includegraphics[width=3.8cm]{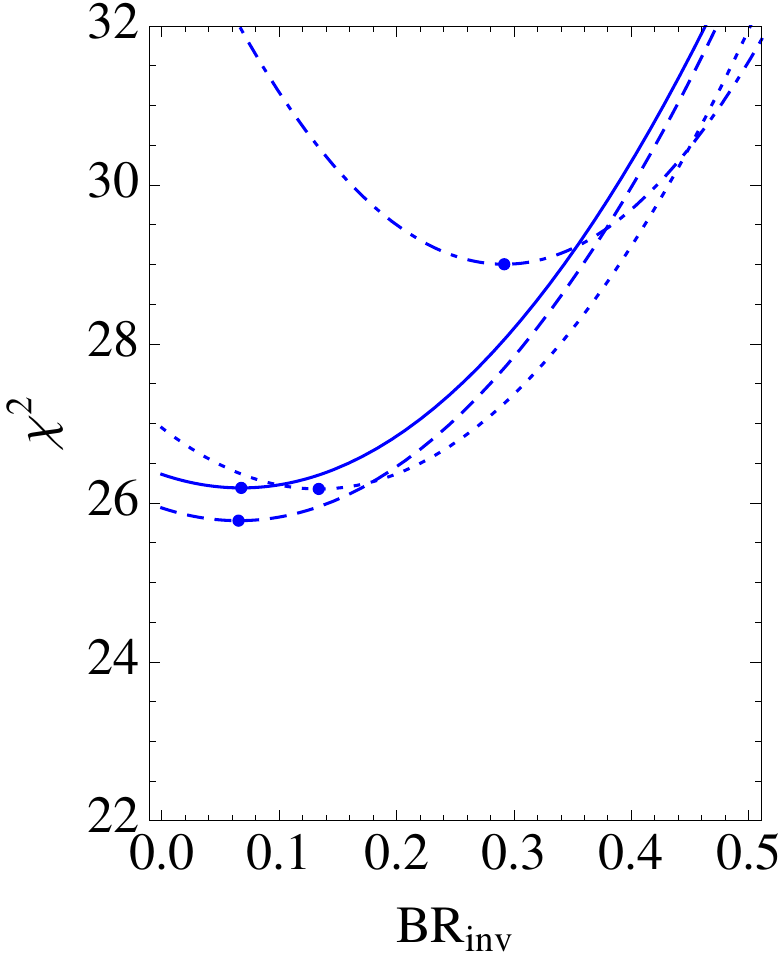}
    \end{minipage}
     \caption{\footnotesize \emph{The $\chi^2$ for invisible branching ratio $BR_{inv}\equiv\Gamma_h^{inv}/\Gamma_h^{SM}$ for different values of $\xi=0, 0.1, 0.25, 0.5$ (solid, dashed, dotted, dot-dashed). The left plot is for $m_h=125$ GeV, the right one uses the CMS data at $m_h=125$ GeV and the ATLAS data at $m_h=126.5$ GeV. Dots correspond to the best fit points.}}\label{figureInv}
\end{figure}

The region where $ \frac{\Gamma_{\eta\rightarrow\gamma\gamma}}{\Gamma_{h\rightarrow\gamma\gamma}}$ vanishes, corresponds to a situation in which all couplings of the Higgs are reduced w.r.t. their SM values by a factor $\cos\alpha$, on top of the suppression due to compositeness. From the point of view of Higgs searches, this is indistinguishable from the situation in which invisible decays reduce the visible branching fraction\footnote{Searches of monojet or dijets plus missing transverse energy can in principle differentiate these possibilities.}, with the identification $\cos^2\alpha \rightarrow \Gamma_h^{vis}/ (\Gamma_h^{vis}+\Gamma_h^{inv}$), where the superscripts differentiate between the total visible and invisible decay width. In the $SO(6)/SO(5)$ models, hidden Higgs decays are possible  in the presence of an extra unbroken $Z_2$ symmetry  that makes $\eta$ stable, and an approximately unbroken $U(1)_\eta$ symmetry that naturally realizes $m_\eta<m_h/2$ \cite{Frigerio:2012uc}. In fig.~\ref{figureInv}, we analyze this  possibility by computing the $\chi^2$ with and without invisible width and for different degrees of compositeness $\xi=0,0.1,0.25,0.5$ (as described in ref.~\cite{Frigerio:2012uc}, in order to realize an approximate $U(1)_\eta$, top and bottom quarks must have different coupling structures to the strong sector, leading to $c_t=\sqrt{1-\xi}$ and $c_b=(1-2\xi)/\sqrt{1-\xi}$; such non-universal couplings of the SM fermions with $h$ lead  to a bottom/tau-phobic top-philic $h$ as $\xi\rightarrow 1/2$, a value however disfavored by EWPD). The limit $\xi\rightarrow 0$, corresponds to the SM with invisible decays (see also \cite{Giardino:2012ww,Espinosa:2012vu}).
Although the best fit point corresponds to $Br_{inv}\neq0$, the $\chi^2$ is rather flat and $Br_{inv}=0$ is in the region preferred by the present data. For larger $\xi$, the preference for $BR_{inv}\neq0$ is stronger. Furthermore, the situation with a small degree of compositeness $\xi$ is slightly better than the SM, as shown by the dashed curves in fig.~\ref{figureInv}. This can be readily understood from fig.~\ref{figureghff}: the SM with an increasing invisible decay width corresponds to moving along the line $n=0$ towards the origin, trajectories where $c$ is suppressed faster than $a$ are preferred by the data.

%%%%%%%%%%%%%%%%%%%%%%%%%%%%%%%%%%%%%%%%%%%%%%%%%%%%%%%%%%%%
%%%%%%%%%%%%%%%%%%%%%%%%%%%%%%%%%%%%%%%%%%%%%%%%%%%%%%%%%%%%
\subsection{Higgs impostors?}

Independently from the composite Higgs realization, the fact that $h$ can mix with other states that do not necessarily participate in the breaking of the electroweak symmetry (an analog example is the two Higgs doublet model discussed in the next section), raises the question of whether the state observed at the LHC is or is not the one whose VEV generates $m_W$ and $m_Z$. This can be done by measuring the parameter $a$ independently: while the Higgs field can have trilinear renormalizable couplings to $WW$ and $a=1$, impostors will have to couple to $WW$ via loops (also the dilaton would couple to matter as $m/f$ and could reproduce the observed excesses~\cite{Ellis:2012rx,Cheung:2011nv}; it is however unlikely that, if $f\approx v$, there would have not been other observable deviations from the SM~\cite{Vecchi:2010gj}). Therefore $a<1$ might imply that the state we observe is not \emph{the} Higgs, or that it is a Higgs that mixes with another state.

\begin{figure}[h]
\centering
   \includegraphics[width=6cm]{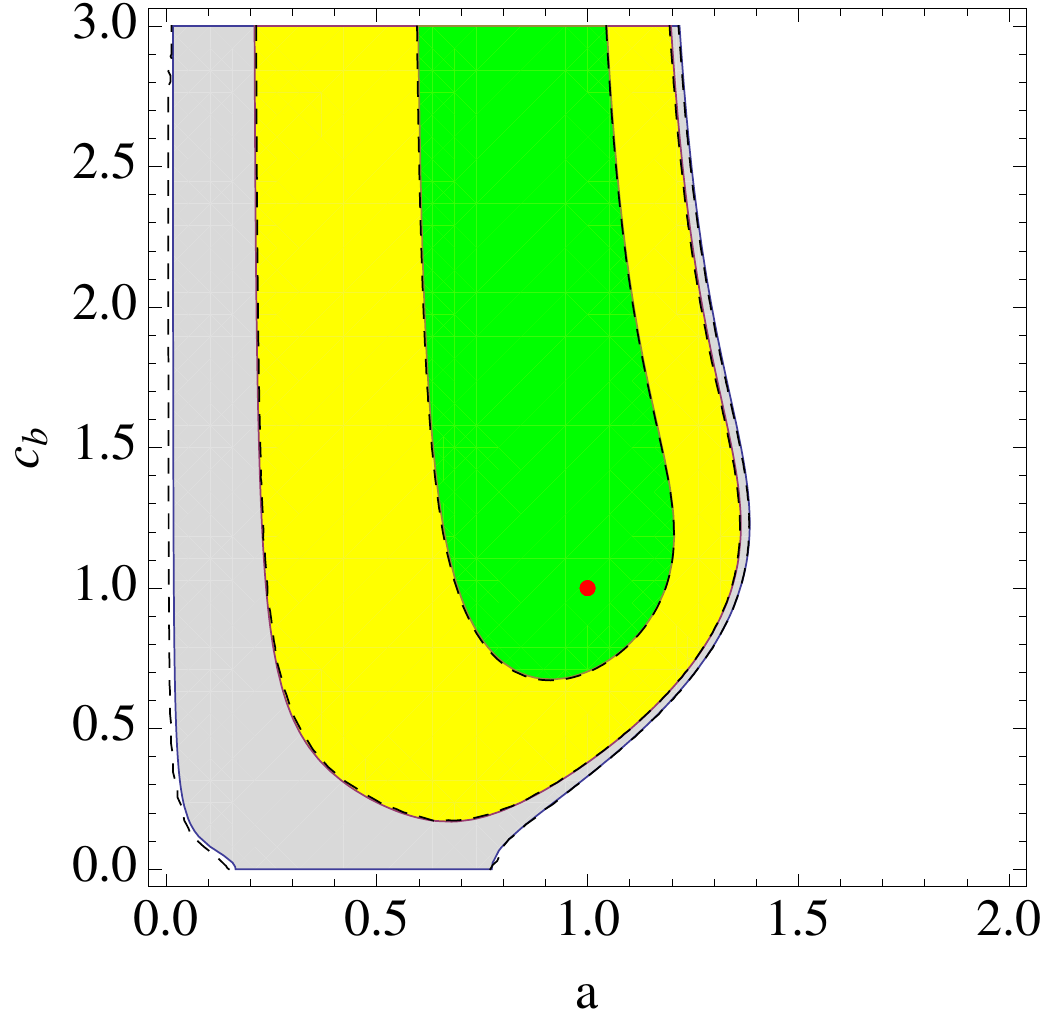}
 \caption{\footnotesize \emph{Preferred regions in the plane $(a,c_b)$ using only the exclusive channels with associated  production (CMS and Tevatron) and VBF cuts (CMS). Colors as in fig.~1. A flat prior is assumed for $a\in[0,3]$ and $c_b\in[0,3]$ (this choice for the upper limit in $c_b$ leads to the most conservative conclusions).}}
 \label{figureVBF}
\end{figure}

One possibility to answer this question, is to marginalize over all parameters except $a$%( see for instance ref.~\cite{Low:2012rj})
. Since we don't know whether some of the Higgs couplings have large deviations from the SM values, another possibility is to isolate some channels that are mostly sensible to $a$ and are sensible to the least number of other parameters. In particular we choose the exclusive VBF channels $pp\rightarrow hjj\rightarrow WW jj$ measured by CMS \cite{Collaboration:2012tx008}, which scales roughly as $\sim a^4/c_b^2$  and the exclusive associated production $pp\rightarrow V h\rightarrow V\bar{b}b$ measured both at Tevatron \cite{:2012cn}, CMS and ATLAS, which scales roughly as $\sim a^2$. These channels are mostly insensitive to $c_t$ and $c_\tau$ and allow a study of $a$ with $c_b$ as only other parameter. Fig.~\ref{figureVBF} shows that, unless $c_b\ll 1$, values close to $a\approx 1$ are preferred by data (we have checked that the influence of the other parameters is negligible).

%%%%%%%%%%%%%%%%%%%%%%%%%%%%%%%%%%%%%%%%%%%%%%%%%%%%%%%%%%%%
%%%%%%%%%%%%%%%%%%%%%%%%%%%%%%%%%%%%%%%%%%%%%%%%%%%%%%%%%%%%
\section{$SO(6)/SO(4)\times SO(2)$ and natural two Higgs doublets models}\label{sec:642}
%%%%%%%%%%%%%%%%%%%%%%%%%%%%%%%%%%%%%%%%%%%%%%%%%%%%%%%%%%%%
%%%%%%%%%%%%%%%%%%%%%%%%%%%%%%%%%%%%%%%%%%%%%%%%%%%%%%%%%%%%

\begin{figure}[h]
\centering
   \includegraphics[width=7cm]{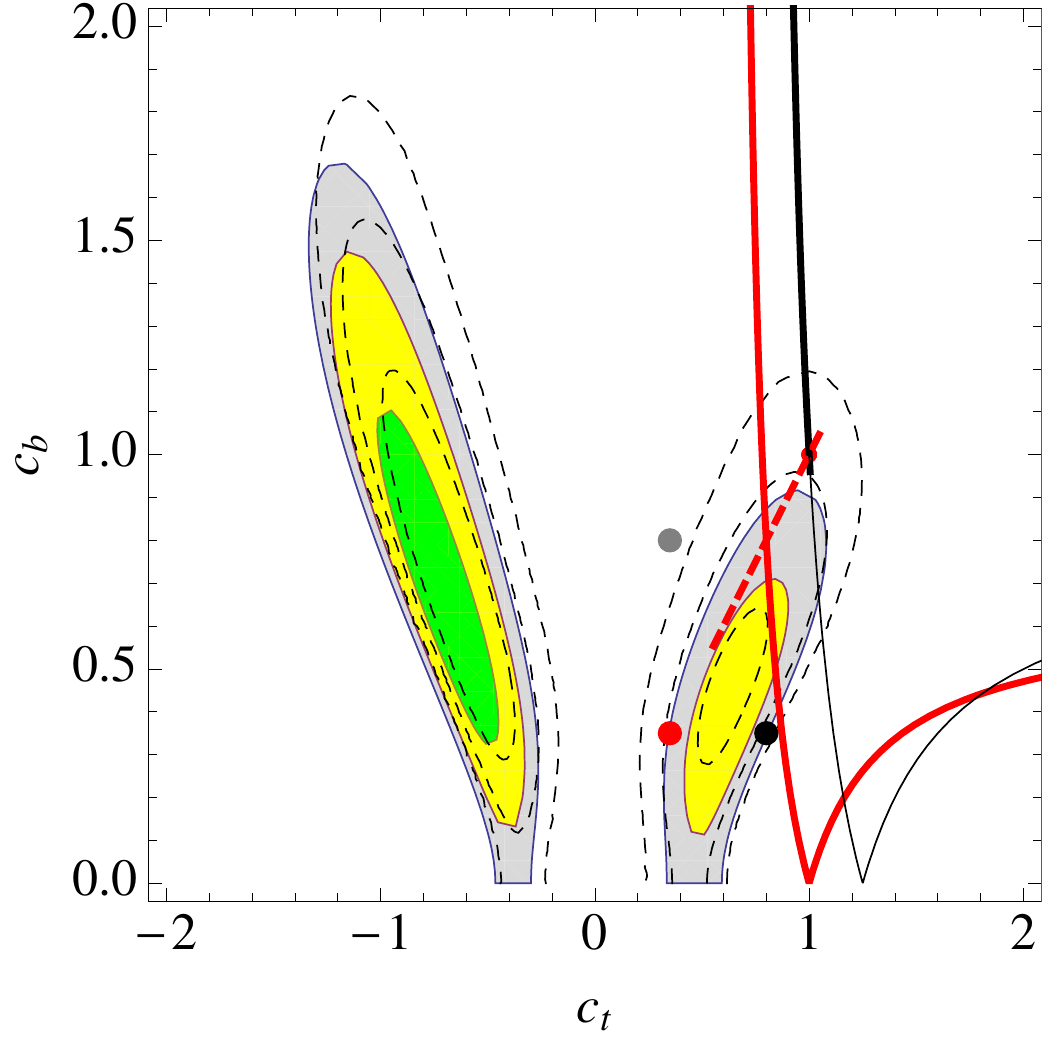}
      \includegraphics[width=7cm]{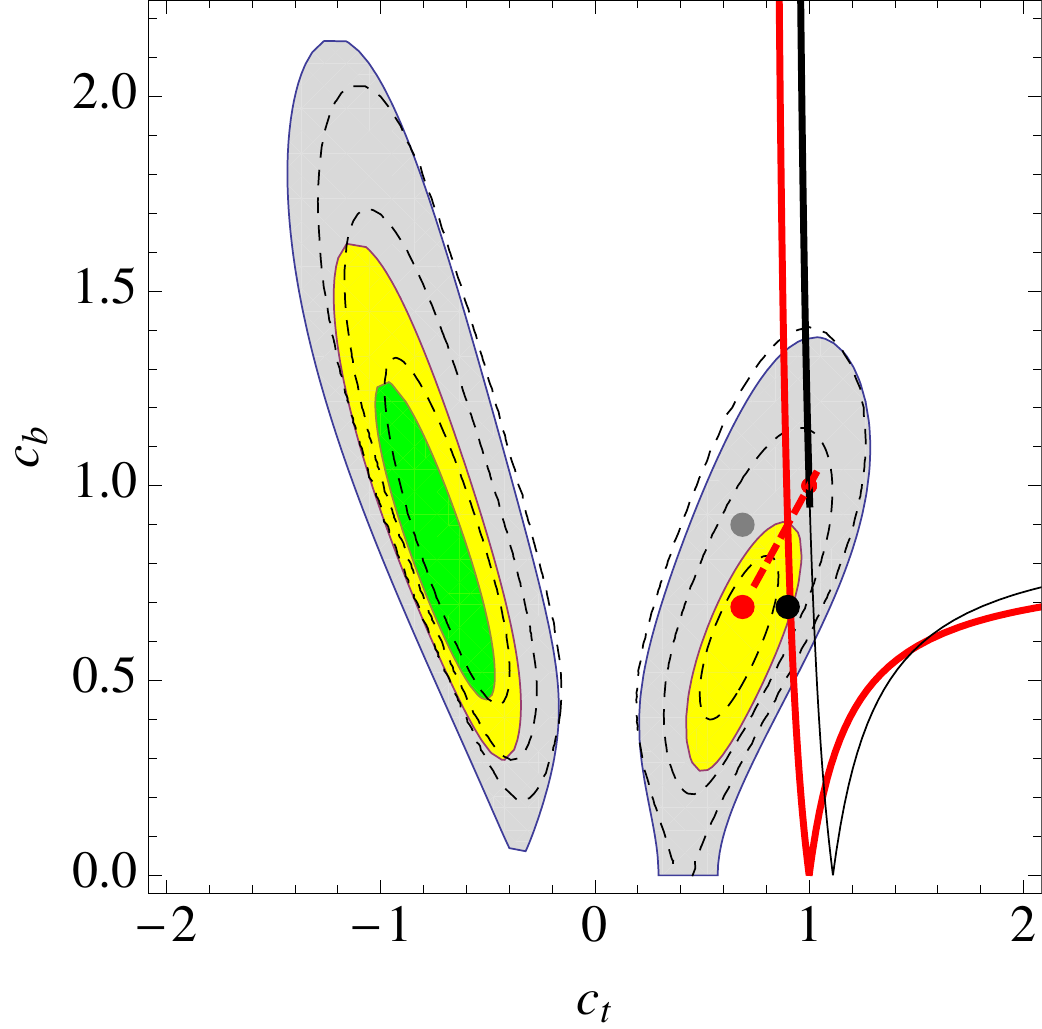}
 \caption{\footnotesize \emph{C.L. contours (colors as fig.~1) in the ($c_t$,$|c_b|$)-plane for $c_b=c_\tau$ and fixed values of $a$=0.8, 0.9  (upper and lower plot respectively). The black line corresponds to the couplings of the elementary (composite)  THDM \eq{alfabeta}, once fixed $a$=0.8, 0.9; the thick part is accessible to the tree-level MSSM. Red lines are for the composite THDM Type II (solid) and Type III (red dashed, varying $-\xi\lesssim\alpha\lesssim \xi$). Shown are also two points (black and gray) corresponding to the SO(5)/SO(4) coset, with $c_t=\sqrt{1-\xi}$ and $c_b=(1-2\xi)/\sqrt{1-\xi}$ and vice versa; the red point is the MCHM5 for comparison. A flat prior $c_t\in[-4,4]$ and $c_b\in[0,4]$ is used.
 }}
 \label{figureTHDM1}
\end{figure}

The $SO(6)/SO(4)\times SO(2)$ coset delivers at low energy 8 PNGBs that can be identified with an effective THDM \cite{Mrazek:2011iu}. Large contributions to the $\hat{T}$-parameter can be avoided thanks to the symmetry $C_2: (H_1,H_2)\rightarrow(H_1,-H_2)$, which also allows us to differentiate three cases:
\begin{itemize}
 \item[\emph{i)}] If $C_2$ is exact, the model is a Type I THDM. The second doublet is heavy and inert \cite{Barbieri:2006dq} and there is no mixing between the CP-even states; this model resembles the SO(5)/SO(4) models of section \ref{sec:54}.\footnote{A similar $h_0$ phenomenology is realized in the \emph{almost inert} model of Ref.~\cite{Mrazek:2011iu}, where the $C_2$ symmetry is unbroken by the top-quark couplings and is broken only by the smaller Yukawas.}
 %%%%%%%%%%%%%%%%%%%%%%%%%%%%%%%%%%%%
 \item[\emph{ii)}] If $C_2$ is spontaneously broken, an effective Type II THDM is realized at low energy, but only at the price of large fine-tuning (both Higgs VEVs have to be tuned much smaller than $f$). The couplings of a Type II THDM are \cite{Gunion:1989we}
\begin{equation}\label{alfabeta}
\frac{y_t^{THDM}}{y_t}=\frac{\cos \alpha}{\sin \beta},\,\,\, \frac{y_b^{THDM}}{y_b}=-\frac{\sin \alpha}{\cos \beta},
\end{equation}
where $\tan\beta=v_1/v_2$ is the ratio between VEVs, and $\alpha$ is the mixing angle (the analog of \eq{Mixing} but for two doublets); the coupling of the lightest Higgs to the vectors is reduced, in comparison with the case of only one Higgs doublet, by $\sin(\beta-\alpha)$.
On top of mixing, also higher dimension operators reduce the couplings between $h$ and the SM vectors and fermions $f$ (as in all composite Higgs models) and for small $\xi$ we obtain
\begin{eqnarray}
a&=&(1-\xi/2)\sin(\beta-\alpha),\nonumber\\
c_t&=&(1-\xi/2)\frac{\cos \alpha}{\sin \beta},\label{THDMeq}\\
c_{b,\tau}&=&-(1-\xi/2)\frac{\sin \alpha}{\cos \beta}.\nonumber\end{eqnarray}
We compare this model  with the tree-level MSSM\footnote{Notice that if some superpartners  (such as staus \cite{Carena:2011aa} or other states \cite{Blum:2012ii,Bellazzini:2012mh}) are light, the rate $h\rightarrow\gamma\gamma$ can be enhanced and the MSSM fit might change considerably; other loop-effects that contribute to the Higgs quartic, can also change this prediction \cite{Azatov:2012wq}.} in fig.~\ref{figureTHDM1} in the ($c_t,c_b$)-plane with $c_b=c_\tau$ and for different slices of $a=0.8, 0.9$ (see ref.~\cite{Azatov:2012wq} for an approach with $a$ marginalized). The black line shows the prediction for the MSSM varying $\beta$ ($\alpha$ is fixed by the slice choice $a=0.8,0.9$); the thin part of the line is unaccessible to the tree-level MSSM due to the peculiar relation between the quartic couplings in the potential \cite{Gunion:1989we}. The composite THDM is drawn in red.
 %%%%%%%%%%%%%%%%%%%%%%%%%%%%%%%%%%%%
 \item[\emph{iii)}] If $C_2$ is explicitly broken, then a small VEV $\langle H_2 \rangle\neq 0$, leading to $\tan\beta\approx\xi^{-1}$, and a small mixing $\tan\alpha\lesssim \xi$ is generated \cite{Mrazek:2011iu}. In this case a Type III THDM originates, in which both $H_1$ and $H_2$ couple to each SM fermion $f$,
\begin{equation}
\frac{y_f}{\sqrt{2}}\bar{f}f(H_1+a_fH_2)=
\frac{y_f}{\sqrt{2}}\bar{f}f(\cos\alpha+a_f\sin\alpha)h_1^0+\cdots\, ,
\label{TypeIII}
\end{equation}
where we have retained only the interactions of the lightest CP-even state $h_1^0$, and where $y_f=(1-\xi/2)y_f^{SM}$, as discussed above. For FCNC to be suppressed, the Yukawas for $H_{1,2}$ must be aligned \cite{Gunion:1989we}.
In fig.~\ref{figureTHDM1}, red dashed line, we show the situation with $a_t=a_{b,\tau}\approx 1$, varying $-\xi\lesssim\alpha\lesssim \xi$.
\end{itemize}
As shown in fig.~\ref{figureTHDM1}, the preferred region is along the direction $c_t\approx -c_b=-c_\tau$ or, with less significance, along $c_t\approx c_b=c_\tau$. Despite the different possibilities realized in composite THDMs, none touches the first preferred region, and the models that are preferred by the data are those closer to the line $c_t\approx c_\tau \approx c_b=c$, which are more similar, in terms of $h$ phenomenology, to the MCHM with $SO(5)/SO(4)$.

%%%%%%%%%%%%%%%%%%%%%%%%%%%%%%%%%%%%%%%%%%%%%%%%%%%%%%%%%%%%%%%%%%%%%%%%%%
%%%%%%%%%%%%%%%%%%%%%%%%%%%%%%%%%%%%%%%%%%%%%%%%%%%%%%%%%%%%%%%%%%%%%%%%%%
%%%%%%%%%%%%%%%%%%%%%%%%%%%%%%%%%%%%%%%%%%%%%%%%%%%%%%%%%%%%
%%%%%%%%%%%%%%%%%%%%%%%%%%%%%%%%%%%%%%%%%%%%%%%%%%%%%%%%%%%%
\section{Conclusions}
%%%%%%%%%%%%%%%%%%%%%%%%%%%%%%%%%%%%%%%%%%%%%%%%%%%%%%%%%%%%
%%%%%%%%%%%%%%%%%%%%%%%%%%%%%%%%%%%%%%%%%%%%%%%%%%%%%%%%%%%%
Using global fits with the most recent data provided by ATLAS, CMS and Tevatron, we have analyzed the parameter space given by couplings of the Higgs to top and bottom quarks, taus and vector bosons. We have shown that the hypothesis of a SM Higgs with  $m_h\approx125$ GeV agrees well with the data.

We have then studied different models, in particular in the context of composite Higgs, to see what features could improve/worsen the situation of these scenarios when more data is available. In particular, we have shown that, depending on the coupling structure of elementary fermions to the strong sector, the couplings of $h$ to the SM fields can change considerably w.r.t. the SM case.  Some composite models, such as the MCHM4, seem to point towards the disfavored direction. Other models, however, follow better the trend of data: the MCHM5, for instance, reduces the couplings of $h$ to fermions, $c$, more than the one to vectors, $a$, and for small $\xi$ crosses the best fit region with $c>0$. Models with $n\gtrsim4$ in \eq{deformation} have an even better trend, as the best fit region with $c_t <0$ lies within their parameter space even for $f\gtrsim 500\GeV$. Different couplings of top and bottom quarks to the strong sector, on the other hand do not seems to ameliorate much the fit, although more drastic possibilities could be considered, that reproduce the preferred region $c_t=-c_b=-c_\tau$, shown in fig.~\ref{figureTHDM1}.

We have  also studied  larger coset structures, such as SO(6)/SO(5), which contains an extra singlet in its light spectrum. We have shown that in this case, large invisible decays widths into a stable singlet, start to be in conflict with the data, although a small one cannot be excluded. On the other hand, a situation in which the Higgs mixes with the singlet, which in turn has enhanced anomalous couplings to $\gamma\gamma$, is in good agreement with the data.

Finally we have discussed  $SO(6)/SO(4)\times SO(2)$ models which, in some cases reduce to a composite model with an inert doublet  (Type I THDM) and the light Higgs phenomenology is not much affected, thus resembling  to the SO(5)/SO(4) coset. If more fine-tuning is allowed, a version of a Type II THDM is possible; compared with the MSSM this has the advantage of having less constrained quartics, which allows the model to have both $c_b>1$ and $c_b<1$, thus covering a larger region of parameter space \cite{Azatov:2012wq}. Despite this,  present data show a mild preference for the composite THDM only for sizable deviations from $a=1$ (upper plot of fig.~\ref{figureTHDM1}).  Finally, if FCNC can be kept under control, a version of Type III THDM is also possible. In this case the couplings of the second Higgs doublet to fermions enter as new parameters in the theory; despite this freedom, the greatest overlap with the best fit regions occur along $c_t\approx c_{b,\tau}$ which is the region also touched by the minimal model. In summary, the composite THDM provides its best fit to the data when  the parameter space is such that its phenomenology resembles much that of the minimal composite Higgs model $SO(5)/SO(4)$.

{\bf Note:} While this work was in preparation, refs.~\cite{Low:2012rj,Giardino:2012dp,Corbett:2012dm} appeared, also discussing deviations of Higgs couplings from its SM values.

%\begin{center}
%{\bf Acknowledgments}
%\end{center}
\section*{Acknowledgments}
 We are especially thankful to  A. Pomarol for encouragement and comments on the manuscript and to M. Farina for the continuos help. We also thank  A.~Azatov, J.R.~Espinosa, J.~Galloway, B.~Gripaios, R.~Maffeis, M.~Redi, J.~Serra, M.~Strassler, A.~Urbano and T.~You for useful discussions. We are grateful to the CERN TH division for hospitality during the completion of this work. MM is supported by the Universitat Aut\`onoma de Barcelona PR-404-01-2/08.

\section*{APPENDIX}
\begin{figure}[t]
\centering
      \includegraphics[width=6cm]{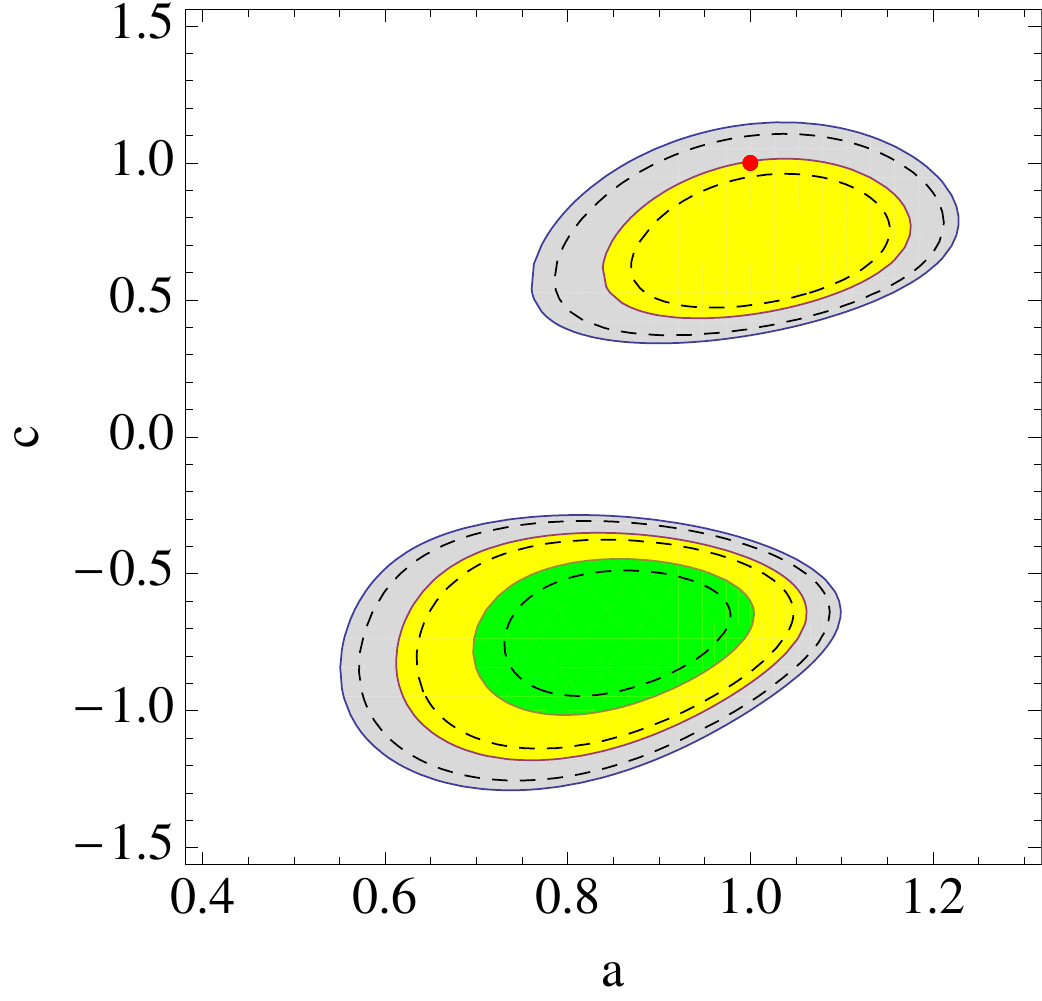}
 \caption{\footnotesize \emph{Comparison between data analyses based on Bayesian interval (as used throughout this work) and isocontours of constant $\chi^2=\chi^2_{min}+ 2.3,5.99,9.21$ (the dashed ones).
 }}
 \label{Figcomparison}
\end{figure}
\noindent
{\bf Bayesian interval versus $\chi^2$ analyses}\\
In fig.~\ref{Figcomparison} we compare the effects of using analyses based on $\chi^2$ (where the point $\chi^2_{min}$ is first found and then the 68\%,95\%,99\%C.L. intervals are found as isocontours with, in the case two fitting parameters, $\chi^2=\chi^2_{min}+ 2.3,5.99,9.21$), versus analyses that use Bayesian intervals \cite{stats} (as done throughout this work). If $\mu$ is just a parameter and not a function depending on $a,c_t,c_b,c_\tau$, the methods do not differ; however, when $\mu=\mu(a,c_t,c_b,c_\tau)$ is a function of the parameters $a,c_{t,b,\tau}$ as in \eq{equation1}, then the probability density function is no longer Gaussian in $a,c_{t,b,\tau}$ and the two methods differ. As shown in fig.~\ref{Figcomparison}, however, the small differences do not alter the qualitative conclusions.

\noindent
{\bf Cut efficiencies $\zeta_p^ i$}\\
The production cross-section for channel $i$ receives contributions from gluon fusion (G), vector boson fusion (VBF) associated production with a vector boson (A) and associated $t\bar{t}$ production ($tth$),
\begin{eqnarray}
\frac{\sum_p\sigma_p\zeta_p^i}{\sum_p\sigma_p^{SM}\zeta_p^i}=&\frac{c_t^2( \, \sigma_{G}\zeta_{G}^i + \,\sigma_{tth}\zeta_{tth}^i)+a^2( \sigma_{VBF}\zeta_{VBF}^i+ \sigma_{A}\zeta_{A}^i)}
{\sigma_{G}\zeta_{G}^i+\sigma_{VBF}\zeta_{VBF}^i+\sigma_{A}\zeta_{A}^i+\sigma_{tth}\zeta_{tth}^i},
\nonumber
\end{eqnarray}
where the cut efficiencies $\zeta_p^i$ for each production mode $p$ corresponding to channel  $i$ from table~\ref{tableChannels} are as follows: when only G, VBF or A is indicated, we have assumed no contamination from other production channels; inclusive channels correspond to $\zeta_{G}^i=\zeta_{VBF}^i=\zeta_{A}^i=\zeta_{tth}^i=1$; other channels, denoted $\gamma\gamma_X$ in table~\ref{tableChannels}, are reported below, where the numbers in brackets denote efficiencies at 8 TeV, the others at 7 TeV \cite{CMSPhotons},
\begin{center}
\begin{tabular}{|c|c|c|c|c|}
\hline
  $i$   & $\zeta_{G}^i$ & $\zeta_{VBF}^i$ & $\zeta_{A}^i$ & $\zeta_{tth}^i$ \\ \hline
 $\gamma\gamma_0$ & 0.28(0.45) & 1(1) &1.52(1.91) & 2.33(4) \\ \hline
 $\gamma\gamma_1$ & 1.16(1.2) & 1(1) &1.36(1.4)&0(0) \\ \hline
 $\gamma\gamma_2$ & 1.82(1.84) & 1(1) &1.36(1.4) &0(0) \\ \hline
 $\gamma\gamma_3$ & 1.82(1.84) & 1(1) &1.36(1.4) &0(0)\\ \hline
  $\gamma\gamma_{jj}$ & 0.029 & 1 &0.01& 0 \\ \hline
  $\gamma\gamma_{jj}$ (T) &(0.024) & (1) &(0)& (0) \\ \hline
  $\gamma\gamma_{jj}$(L) &(0.094) & (1) &(0.063)& (0) \\ \hline
\end{tabular}
\end{center}
and the overall normalization in each line factorizes.


\begin{thebibliography}{99}

\bibitem{CERNTalkCMS}
J. Incandela, CMS talk at \emph{Latest update in the search for the Higgs boson} at CERN, July 4, 2012;

\bibitem{CERNTalkATLAS}
F. Gianotti, ATLAS talk at \emph{Latest update in the search for the Higgs boson} at CERN, July 4, 2012.

%\cite{Agashe:2004rs}
\bibitem{Agashe:2004rs}
  K.~Agashe, R.~Contino and A.~Pomarol,
  %``The Minimal composite Higgs model,''
  Nucl.\ Phys.\ B {\bf 719} (2005) 165.
 % [hep-ph/0412089].
  %%CITATION = HEP-PH/0412089;%%

%%\cite{Contino:2006qr}
\bibitem{Contino:2006qr}
  R.~Contino, L.~Da Rold and A.~Pomarol,
  %``Light custodians in natural composite Higgs models,''
  Phys.\ Rev.\ D {\bf 75} (2007) 055014.
 % [hep-ph/0612048].
  %%CITATION = HEP-PH/0612048;%%

    %\cite{Pomarol:2012qf}
\bibitem{Pomarol:2012qf}
  A.~Pomarol and F.~Riva,
  %``The Composite Higgs and Light Resonance Connection,''
  arXiv:1205.6434 [hep-ph].
  %%CITATION = ARXIV:1205.6434;%%

    %\cite{Gripaios:2009pe}
\bibitem{Gripaios:2009pe}
 B.~Gripaios, A.~Pomarol, F.~Riva, J.~Serra,
 %``\textit{Beyond the Minimal Composite Higgs Model}'',
 JHEP {\bf 0904 } (2009)  070.
 %[hep-ph/0902.1483].
   %%CITATION = ARXIV:0902.1483;%%

     %\cite{Mrazek:2011iu}
\bibitem{Mrazek:2011iu}
  J.~Mrazek, A.~Pomarol, R.~Rattazzi, M.~Redi, J.~Serra and A.~Wulzer,
  %``The Other Natural Two Higgs Doublet Model,''
  Nucl.\ Phys.\ B {\bf 853} (2011) 1.
%  [arXiv:1105.5403 [hep-ph]].
  %%CITATION = ARXIV:1105.5403;%%


  %\cite{Espinosa:2010vn}
\bibitem{Espinosa:2010vn}
  J.~R.~Espinosa, C.~Grojean and M.~Muhlleitner,
  %``Composite Higgs Search at the LHC,''
  JHEP {\bf 1005} (2010) 065
  [arXiv:1003.3251 [hep-ph]].
  %%CITATION = ARXIV:1003.3251;%%

  %\cite{Carmi:2012yp}
\bibitem{Carmi:2012yp}
  D.~Carmi, A.~Falkowski, E.~Kuflik and T.~Volansky,
  %``Interpreting LHC Higgs Results from Natural New Physics Perspective,''
  arXiv:1202.3144 [hep-ph].
  %%CITATION = ARXIV:1202.3144;%%

  %\cite{Azatov:2012bz}
\bibitem{Azatov:2012bz}
  A.~Azatov, R.~Contino and J.~Galloway,
  %``Model-Independent Bounds on a Light Higgs,''
  JHEP {\bf 1204} (2012) 127
  [arXiv:1202.3415 [hep-ph]].
  %%CITATION = ARXIV:1202.3415;%%

  %\cite{Espinosa:2012ir}
\bibitem{Espinosa:2012ir}
  J.~R.~Espinosa, C.~Grojean, M.~Muhlleitner and M.~Trott,
  %``Fingerprinting Higgs Suspects at the LHC,''
  JHEP {\bf 1205} (2012) 097.
  %[arXiv:1202.3697 [hep-ph]].
  %%CITATION = ARXIV:1202.3697;%%


  %\cite{Ellis:2012rx}
\bibitem{Ellis:2012rx}
  J.~Ellis and T.~You,
  %``Global Analysis of Experimental Constraints on a Possible Higgs-Like Particle with Mass ~ 125 GeV,''
  arXiv:1204.0464 [hep-ph].
  %%CITATION = ARXIV:1204.0464;%%

  \bibitem{crossx}
  https://twiki.cern.ch/twiki/bin/view/LHCPhysics/ CrossSections






%    %\cite{Chatrchyan:2012tw}
%\bibitem{Chatrchyan:2012tw}
%  S.~Chatrchyan {\it et al.}  [CMS Collaboration],
%  %``Search for the standard model Higgs boson decaying into two photons in pp collisions at sqrt(s)=7 TeV,''
%  Phys.\ Lett.\ B {\bf 710} (2012) 403
%  [arXiv:1202.1487 [hep-ex]].
%  %%CITATION = ARXIV:1202.1487;%%

%\cite{CMSPhotons}
\bibitem{CMSPhotons}
  CMS Collaboration,
    \href{https://twiki.cern.ch/twiki/bin/view/CMSPublic/Hig12015TWiki}{CMS PAS HIG-12-015}.


%%\cite{Collaboration:2012tx001}
%\bibitem{Collaboration:2012tx001}
%  CMS Collaboration,
%    \href{http://cdsweb.cern.ch/record/1429931/files/HIG-12-001-pas.pdf}{CMS PAS HIG-12-001}.


%\cite{Collaboration:2012tx008}
\bibitem{Collaboration:2012tx008}
  CMS Collaboration,
    \href{http://cdsweb.cern.ch/record/1429928/files/HIG-12-008-pas.pdf}{CMS PAS HIG-12-008}.

%    %\cite{ATLAS:2012tx019}
%\bibitem{ATLAS:2012tx019}
%  ATLAS Collaboration,
%    \href{http://cdsweb.cern.ch/record/1430033/files/ATLAS-CONF-2012-019.pdf}{ATLAS-CONF-2012-019}.

    \bibitem{Atlasphoton}
    Tackmann~K., %``Search for the Higgs boson in the diphoton decay channel with the ATLAS detector,''
    talk at ICHEP 2012.

%\bibitem{ATLAS7}
%  G.~Aad {\it et al.}  [ATLAS Collaboration],
%  %``Combined search for the Standard Model Higgs boson in pp collisions at sqrt(s) = 7 TeV with the ATLAS detector,''
%  arXiv:1207.0319 [hep-ex].
%  %%CITATION = ARXIV:1207.0319;%%

%\cite{ATLAS:2012tx019}
\bibitem{ATLAS:2012tx019}
  ATLAS Collaboration,
    \href{http://cdsweb.cern.ch/record/1430033/files/ATLAS-CONF-2012-019.pdf}{ATLAS-CONF-2012-019}.

%%\cite{Aad:2012sc}
%\bibitem{Aad:2012sc}
%  G.~Aad {\it et al.}  [ATLAS Collaboration],
%  %``Search for the Standard Model Higgs boson in the H -> WW(*) -> l nu l nu decay mode with 4.7 /fb of ATLAS data at sqrt(s) = 7 TeV,''
%  arXiv:1206.0756 [hep-ex].
%  %%CITATION = ARXIV:1206.0756;%%



%\cite{:2012cn}
\bibitem{:2012cn}
 [The CDF Collaboration],
  %``Updated Combination of CDF and D0 Searches for Standard Model Higgs Boson Production with up to 10.0 fb-1 of Data,''
  arXiv:1207.0449 [hep-ex].
  %%CITATION = ARXIV:1207.0449;%%

      %\cite{Giudice:2007fh}
\bibitem{Giudice:2007fh}
  G.~F.~Giudice, C.~Grojean, A.~Pomarol and R.~Rattazzi,
  %``The Strongly-Interacting Light Higgs,''
  JHEP {\bf 0706} (2007) 045
  [hep-ph/0703164].
  %%CITATION = HEP-PH/0703164;%%

   %\cite{Frigerio:2012uc}
\bibitem{Frigerio:2012uc}
  M.~Frigerio, A.~Pomarol, F.~Riva and A.~Urbano,
  %``Composite Scalar Dark Matter,''
  arXiv:1204.2808 [hep-ph].
  %%CITATION = ARXIV:1204.2808;%%

  %\cite{Falkowski:2007hz}
\bibitem{Falkowski:2007hz}
  A.~Falkowski,
  %``\textit{Pseudo-goldstone Higgs production via gluon fusion}'',
  Phys.\ Rev.\ D {\bf 77}, 055018 (2008);
%  [arXiv:0711.0828 [hep-ph]]
  %%CITATION = ARXIV:0711.0828;%%
  I.~Low and A.~Vichi,
  %``\textit{On the production of a composite Higgs boson}'',
  Phys.\ Rev.\ D {\bf 84}, 045019 (2011);
 % [arXiv:1010.2753 [hep-ph]].
  %%CITATION = ARXIV:1010.2753;%%

  % \cite{Azatov:2011qy}
\bibitem{Azatov:2011qy}
  A.~Azatov and J.~Galloway,
  %``\textit{Light Custodians and Higgs Physics in Composite Models}'',
  arXiv:1110.5646 [hep-ph].
  %%CITATION = ARXIV:1110.5646;%%

%\cite{Bellazzini:2012tv}
\bibitem{Bellazzini:2012tv}
  B.~Bellazzini, C.~Csaki, J.~Hubisz, J.~Serra and J.~Terning,
  %``Composite Higgs Sketch,''
  arXiv:1205.4032 [hep-ph].
  %%CITATION = ARXIV:1205.4032;%%



%\cite{Giardino:2012ww}
\bibitem{Giardino:2012ww}
  P.~P.~Giardino, K.~Kannike, M.~Raidal and A.~Strumia,
  %``Reconstructing Higgs boson properties from the LHC and Tevatron data,''
  JHEP {\bf 1206}, 117 (2012).
 % [arXiv:1203.4254 [hep-ph]].
  %%CITATION = ARXIV:1203.4254;%%

    %\cite{Espinosa:2012vu}
\bibitem{Espinosa:2012vu}
  J.~R.~Espinosa, M.~Muhlleitner, C.~Grojean and M.~Trott,
  %``Probing for Invisible Higgs Decays with Global Fits,''
  arXiv:1205.6790 [hep-ph].
  %%CITATION = ARXIV:1205.6790;%%

  %\cite{Cheung:2011nv}
\bibitem{Cheung:2011nv}
  K.~Cheung and T.~-C.~Yuan,
  %``Could the excess seen at 124-126 GeV be due to the Randall-Sundrum Radion?,''
  Phys.\ Rev.\ Lett.\  {\bf 108} (2012) 141602
  [arXiv:1112.4146 [hep-ph]].
  %%CITATION = ARXIV:1112.4146;%%

%\cite{Vecchi:2010gj}
\bibitem{Vecchi:2010gj}
  L.~Vecchi,
  %``Phenomenology of a light scalar: the dilaton,''
  Phys.\ Rev.\ D {\bf 82} (2010) 076009
  [arXiv:1002.1721 [hep-ph]].
  %%CITATION = ARXIV:1002.1721;%%

  %\cite{Low:2012rj}
\bibitem{Low:2012rj}
  I.~Low, J.~Lykken and G.~Shaughnessy,
  %``Have We Observed the Higgs (Imposter)?,''
  arXiv:1207.1093 [hep-ph].
  %%CITATION = ARXIV:1207.1093;%%

  %\cite{Barbieri:2006dq}
\bibitem{Barbieri:2006dq}
  R.~Barbieri, L.~J.~Hall and V.~S.~Rychkov,
  %``Improved naturalness with a heavy Higgs: An Alternative road to LHC physics,''
  Phys.\ Rev.\ D {\bf 74} (2006) 015007
  [hep-ph/0603188].
  %%CITATION = HEP-PH/0603188;%%

  %\cite{Gunion:1989we}
\bibitem{Gunion:1989we}
  For a review, see J.~F.~Gunion, H.~E.~Haber, G.~L.~Kane and S.~Dawson,
  %``The Higgs Hunter's Guide,''
  Front.\ Phys.\  {\bf 80} (2000)~1,
  %%CITATION = FRPHA,80,1;%%
  and
  A.~Djouadi,
  %``The Anatomy of electro-weak symmetry breaking. II. The Higgs bosons in the minimal supersymmetric model,''
  Phys.\ Rept.\  {\bf 459} (2008) 1
  [hep-ph/0503173].
  %%CITATION = HEP-PH/0503173;%%



  %\cite{Arvanitaki:2011ck}
\bibitem{Arvanitaki:2011ck}
  A.~Arvanitaki and G.~Villadoro,
  %``A Non Standard Model Higgs at the LHC as a Sign of Naturalness,''
  JHEP {\bf 1202} (2012) 144
  [arXiv:1112.4835 [hep-ph]].
  %%CITATION = ARXIV:1112.4835;%%

  %\cite{Azatov:2012wq}
\bibitem{Azatov:2012wq}
  A.~Azatov, S.~Chang, N.~Craig and J.~Galloway,
  %``Early Higgs Hints for Non-Minimal Supersymmetry,''
  arXiv:1206.1058 [hep-ph].
  %%CITATION = ARXIV:1206.1058;%%

%\cite{Carena:2011aa}
\bibitem{Carena:2011aa}
  M.~Carena, S.~Gori, N.~R.~Shah and C.~E.~M.~Wagner,
  %``A 125 GeV SM-like Higgs in the MSSM and the $\gamma \gamma$ rate,''
  JHEP {\bf 1203} (2012) 014
  [arXiv:1112.3336 [hep-ph]].
  %%CITATION = ARXIV:1112.3336;%%



  %\cite{Blum:2012ii}
\bibitem{Blum:2012ii}
  K.~Blum, R.~T.~D'Agnolo and J.~Fan,
  %``Natural SUSY Predicts: Higgs Couplings,''
  arXiv:1206.5303 [hep-ph].
  %%CITATION = ARXIV:1206.5303;%%

%\cite{Bellazzini:2012mh}
\bibitem{Bellazzini:2012mh}
  B.~Bellazzini, C.~Petersson and R.~Torre,
  %``Photophilic Higgs from sgoldstino mixing,''
  arXiv:1207.0803 [hep-ph].
  %%CITATION = ARXIV:1207.0803;%%

  %\cite{Giardino:2012dp}
\bibitem{Giardino:2012dp}
  P.~P.~Giardino, K.~Kannike, M.~Raidal and A.~Strumia,
  %``Is the resonance at 125 GeV the Higgs boson?,''
  arXiv:1207.1347 [hep-ph].
  %%CITATION = ARXIV:1207.1347;%%

  %\cite{Corbett:2012dm}
\bibitem{Corbett:2012dm}
  T.~Corbett, O.~J.~P.~Eboli, J.~Gonzalez-Fraile and M.~C.~Gonzalez-Garcia,
  %``Constraining anomalous Higgs interactions,''
  arXiv:1207.1344 [hep-ph].
  %%CITATION = ARXIV:1207.1344;%%

%\cite{stats}
\bibitem{stats}
J.~Beringer et al.(PDG), PR {\bf D86}, 010001 (2012) (http://pdg.lbl.gov)


\end{thebibliography}
\end{document}